\documentclass[journal=jctcce,manuscript=article]{achemso}

\usepackage{amsmath}
\usepackage{mathtools}
\usepackage{braket}
\usepackage{multirow}
\usepackage{graphicx}
\usepackage{times}
\usepackage[table]{xcolor}

\author{Wai-Pan Ng$^\dagger$}
\affiliation[]
{Department of Chemistry, The University of Hong Kong, 
Hong Kong, P.R. China}
\alsoaffiliation[]
{Hong Kong Quantum AI Lab Limited, Hong Kong, P.R. China}
\author{Qiujiang Liang$^\dagger$}
\affiliation[]
{Department of Chemistry, The University of Hong Kong, 
Hong Kong, P.R. China}
\author{Jun Yang}
\affiliation[]
{Department of Chemistry, The University of Hong Kong, 
Hong Kong, P.R. China}
\alsoaffiliation[]
{Hong Kong Quantum AI Lab Limited, Hong Kong, P.R. China}
\email{juny@hku.hk}

\title[]{Low-data deep quantum chemical learning for accurate MP2 and coupled-cluster correlations
\footnote{W.-P. N. and Q. L. contributed equally to this study.}}

\graphicspath{{.}}

\begin{document}


\begin{abstract}
Accurate \textit{ab-initio} prediction of electronic energies is very expensive for macromolecules by explicitly solving post-Hartree-Fock equations. We here exploit the physically justified local correlation feature in compact basis of small molecules, and construct an expressive low-data deep neural network (dNN) model to obtain machine-learned electron correlation energies on par with MP2 and CCSD levels of theory for more complex molecules and different datasets that are not represented in the training set. We show that our dNN-powered model is data efficient and makes highly transferable prediction across alkanes of various lengths, organic molecules with non-covalent and biomolecular interactions, as well as water clusters of different sizes and morphologies. In particular, by training 800 (H$_2$O)$_{8}$ clusters with the local correlation descriptors,  accurate MP2/cc-pVTZ correlation energies up to (H$_2$O)$_{128}$ can be predicted with a small random error within chemical accuracy from exact values, while a majority of prediction deviations are attributed to an intrinsically systematic error. Our results reveal that an extremely compact local correlation feature set, which is poor for any direct post-Hartree-Fock calculations, has however a prominent advantage in reserving important electron correlation patterns for making accurate transferable predictions across distinct molecular compositions, bond types and geometries.
\end{abstract}

\section{Introduction}\label{sec1}

Post-Hartree-Fock (post-HF) quantum chemistry provides an attractive tool for studying molecular chemistry and materials sciences due to its systematically manageable accuracy by directly solving an approximate time-independent Schr\"odinger equation for correlated molecular wave functions. Recent advances in low-scaling post-HF algorithms and implementations have achieved an unprecedented acceleration in computing increasingly complex systems and processes.\cite{yang2014ab, liang2023water} These techniques reduce the computational complexities by commonly adopting the low-dimensional structure of the target wave function in various highly compact and scalable representations that have been  physically inspired by the locality of electron interactions. \cite{pulay1983localizability, saebo1985local}
However, for accurately treating large molecules at a high level of theory, these computations are yet short enough to provide fast solutions, in part due to a significant timing prefactor in the formal scaling, preventing the practical utility from accounting for possible molecular compositions and conformations that increase combinatorially with the number of atoms. The resulting accuracy of these methods is also very sensitive to the approximate level of the inclusion of important electronic configurations.

Recently, a plethora of machine learning (ML) approaches have been developed to aim for a cheap surrogate model with the prediction accuracy of the reference data. A popular end-to-end mapping between the density and potential or density and energy has been established using atomic positions and nuclear charges, \cite{brockherde2017bypassing, tsubaki2021quantum} or converged mean-field solutions, \cite{margraf2021pure, dick2020machine} essentially a reminiscent of the exact density functional theory. Based on the atomistic locality for interatomic interactions, the atomistic ML algorithms for quantum chemistry have been shown to exhibit an astonishing ability in expressing, learning and predicting high dimensional data structure with complex hidden patterns for a variety of target properties if atomic descriptors are handcrafted carefully. Different atomistic ML models have been proposed to predict the total molecular energy by embedding additive atomic contributions which can be learned separately for both kernel methods \cite{rupp2012fast, bartok2013representing, hansen2015machine, faber2018alchemical} and neural networks (NNs). \cite{behler2007generalized, schutt2017quantum, zhang2018deep, schutt2018schnet, grisafi2018transferable} Physically augmented many-body descriptors have been also attempted for including nonadditive interatomic interactions \cite{pronobis2018many, unke2019physnet}.  Meanwhile, the decomposition of the total energy into atomic contributions was lifted through an energy conserving model \cite{chmiela2017machine} or message passing neural networks \cite{gilmer2017neural} for improving  accuracy.

Apart from the computational scaling with respect to atomistic ML parameters and data sizes, there is an atomistic formal complexity of $\mathcal{O}(N^2)$ with respect to the number of atoms ($N$), at the costs similar to classical force field computation if the global pairwise interactions are needed among all atoms to build the accurate local environment, for example, for producing smooth potential energy surfaces with all atomic degrees of freedom. Although it is difficult to determine the \textit{prior} important interatomic interactions, the local atomistic $\mathcal{O}(N)$ ML models can be implemented by neglecting long-range interatomic interactions through a predefined real-space cutoff radius \cite{behler2007generalized, schutt2017quantum, schutt2018schnet, unke2018reactive, grisafi2018transferable} or by augmenting molecular fragments in the training set on the fly. \cite{huang2020quantum} However, these approaches also suggest that the energy prediction of large molecular structures is typically poor if the molecules or molecular fragments that are not sufficiently large are trained to accurately represent the target structure and its chemical environment. On the other hand, the generation of reference dataset containing large molecules is unfeasible by exact quantum chemistry computations, such as coupled-cluster (CC) even on a small dataset (e.g., a few hundred structures), and the number of geometric parameters must increase rapidly with $N$ which leads to an explosive pool of atomic descriptors for training and testing.

Alternative to the atomistic locality,  the electronic locality for explicit electron interactions has been also investigated as a critical element for developing electronic ML methods which represent molecules and electron interactions by learning inexpensive \textit{ab-initio} electronic structure data from, for example, the correlation electron densities \cite{nudejima2019machine, han2021machine}, local electronic orbitals from low-tier wave functions, \cite{welborn2018transferability, cheng2019universal, cheng2019regression, chen2020ground, husch2021improved, cheng2022UC, qiao2020orbnet, qiao2022informing, karandashev2022orbital} post-HF wave function amplitude and density tensors, \cite{margraf2018making, townsend2019data, peyton2020machine, townsend2020transferable} small multireference calculations \cite{king2021ranked} and many others. In principle, the local electronic degrees of freedom are usually near-sighted on a compact subset of atoms and transferable between molecules of different sizes. \cite{meyer2016libraries} The low-tier intermediate electronic descriptors are apparently more expressive than atomistic ones by respecting the quantum mechanically correlated nature of electrons which encodes the part of the mapping from an atomistic geometry to an electronic distribution. {Notably, the recent OrbNet and OrbNet-Equi models \cite{qiao2020orbnet, qiao2022informing} have expressed mean-field operators in local symmetry-adapted atomic orbitals from tight-binding simulations for an embedded graph neural network.} Nevertheless, the electronic locality that had been intensively developed to accelerate direct low-scaling post-HF computations {and its interplay with correlated feature design} have yet been fully explored to further enhance the ML data and learning efficiencies of electronic ML approaches that allow accurate and transferable large-scale predictions.

In this contribution, we introduce a transferable deep NN (T-dNN) model for predicting the energy of a large molecule by learning cheap local wave function operators of small molecules, inspired by the idea that the extremely crude local structure of post-HF wave functions, which is poor for any useful direct post-HF computation, is however systematically improvable, intrinsically encoded with the full  symmetry constraint and can be generated and learned at low costs comparable to the preceding HF computation. More specifically, these coarse-grained wave function operators can be utilized to effectively represent the various essential electron interactions through an electronic decomposition of local descriptors to map the energies of molecules. Our T-dNN model leverages the learning ability of the deep NN to fine-tune and improve the crude mapping along quantum mechanically justified curvature. In this work, we will demonstrate the efficient transferability of our T-dNN model in accurately predicting MP2 and CCSD correlation energies across the organic compound space of various elements, sizes, covalent and non-covalent interactions without the explicit training of large target structures. In particular, we will show its excellent performance in extrapolating the energies of water clusters of various sizes, morphologies and molecular dynamics trajectories up to (H$_2$O)$_{128}$ by training small clusters of (H$_2$O)$_{8}$ and (H$_2$O)$_{16}$. 

\section{Methods}

\subsection{Local Representation of Electronic Descriptors}\label{sec2}

In our approach, the baseline Hartree-Fock (HF) calculations are needed for generating molecular orbitals (MOs). Other calculations which produce MOs, including low-cost self-consistent field approach, \cite{salek2007linear} fragment MO methods \cite{gordon2009accurate} and supervised NN learning scheme, \cite{schutt2019unifying, gastegger2020deep} can be interfaced with the present implementation. While MOs contain the mean-field quantum mechanical feature, the MO-based two-electron features, such as Coulomb integrals, are expressive for mapping a correlated molecular wave function. However, the generation of these electronic descriptors for large molecules is rather computationally demanding, and thus a trade-off between expressivity and complexity is necessary for developing an inexpensive electronic ML approach.  We realize that a crude  electronic structure, which may yield poor electron correlations in direct calculations, can be still topologically similar to the exact solution, if the correlation feature is properly extracted by being able to systematically tune a correlation space in which electronic descriptors are expressed. Therefore, the resulting electronic descriptors can be better expressive and constructed at a low computational expense for improving data and learning efficiencies. 

We express the electronic descriptors in the compact basis of orbital-specific-virtuals (OSVs) correlation space\cite{yang2011tensor, yang2012orbital, schutz2013orbital, zhou2019complete, liang2021third} that is tailored for each occupied local MO (LMO). The OSV correlation space is formed by the orthonormal singular vectors $\mathbf{Q}_{i}$  of the semi-canonical MP2 diagonal pair amplitudes $\mathbf{T}_{ii}=(i a\rvert i b)/\left(\epsilon_{a}+\epsilon_{b}-2f_{ii}\right)$ for each LMO $i$, where  $f_{ii}$ is the diagonal element of the Fock matrix in the localized space,  $\epsilon_a$ (and $\epsilon_b$) the canonical orbital energy of the virtual orbital $a$ (and $b$), and $(i a\rvert i b)$ the Coulomb repulsion integral. {The singular vectors $\mathbf{Q}_{i}$ re-parameterizes the first-order interacting subspace into a more compact representation associated with each occupied LMO $i$, which allows its truncation according to the magnitude of the singular values. Those $\mathbf{Q}_{i}$ with small singular values contribute little to electron correlations and can be neglected. Consequently it is reasonable to collect only those vectors with the largest eigenvalues for which the singular vectors $\mathbf{Q}_{i}$ constitute the majority of the first-order interacting subspace for recovering a large amount of correlations.} The resulting OSV basis vectors $\{\bar\mu_{i}\}$ are expressed as a unitary transformation via selected important $\mathbf{Q}_{i}$ from all canonical virtual orbitals.  In the context of ML, the selection of learning features is automatically prioritized by using the most dominant OSVs according to the singular values, and no feature selection is needed in this approach. We have found that only a few OSVs with the most important singular values are sufficient to build very sparse electronic descriptors to refactor the learning pattern in the OSV basis in which the critical correlation environment is compressed in the vicinity of the LMO for an electron. 
The cost of computing OSVs formally scales as $\mathcal{O}(ON^3)$ with respect to the numbers of occupied LMOs ($O$) and atoms ($N$), and has been substantially reduced to practical $\mathcal{O}(ON)$ by the sparse fitting and interpolative decomposition techniques of our previous implementation,\cite{liang2021third} much faster than HF ($\mathcal{O}(N^4)$). The   generation of OSVs is further significantly accelerated in the parallel computation by uniformly distributing LMOs over all processors. For example, on 24 CPU cores, selected OSVs are generated within only a few seconds for Nonactin  (C$_{40}$H$_{64}$O$_{12}$) with def2-TZVP basis set and only 22.5 seconds for (H$_2$O)$_{190}$ with cc-pVTZ basis set. \cite{liang2021third} 

\subsection{Feature Implementation}

Four feature vectors are adopted to express the correlation energy, including the Fock  $f_{ij}$, Coulomb $J_{ij}$ and exchange $K_{ij}$ elements in the occupied LMOs ($i, j, k, \cdots$), {as well as the contravariant amplitudes $\mathbf{\widetilde{T}}^c_{ij}=2\mathbf{\widetilde{T}}_{ij}$-$\mathbf{\widetilde{T}}_{ij}^{\dagger}$ in the OSV basis, where the superscript $c$ denotes the contravariance. The $\mathbf{\widetilde{T}}^c_{ij}$ is further contracted with the OSV-based exchange integral to generate the input pseudo-energy tensor.} Although the Coulomb and exchange integrals involving virtual orbitals ($a,b,c,\cdots$)  have been used to feature the electron correlation in the previous report,\cite{welborn2018transferability} their computation and sorting are rather expensive. To reduce the cost of computing these two-electron descriptors, we represent the electrostatic repulsion and exchange interactions by only two {occupied-occupied integrals through $J_{ij}=\left(ii\vert jj\right)$ and $K_{ij}=\left(ij\vert ij\right)$ elements, respectively}. They are readily available by transforming baseline HF integrals, and the number of them increases according to $\mathcal{O}(N)$ by neglecting asymptotically unimportant distant $ij$ pairs, {as discussed in Sec. 2.4.} 

We further design feature sets expressed in the OSV basis to physically distinguish the unique electron correlation patterns associated with their first-order interacting subspaces of different characters, 
\begin{equation}
\{\Psi^{(\mathrm{vt})}_{ij},\Psi^{(\mathrm{ex})}_{ij},\Psi^{(\mathrm{ct1})}_{ij},\Psi^{(\mathrm{ct2})}_{ij}\}
\end{equation}
{for an off-diagonal electron pair $ij$ in different LMOs, 
where $\Psi^{(\mathrm{vt})}_{ij}$, $\Psi^{(\mathrm{ex})}_{ij}$,  $\Psi^{(\mathrm{ct1})}_{ij}$ and $\Psi^{(\mathrm{ct2})}_{ij}$ are the compact subspaces encompassing different particle-hole double excitation manifolds: the vertical ($\tilde{t}_{ij}^{\bar\mu_i\bar\nu_j}$: $i \rightarrow \{\bar\mu_i\},~j \rightarrow \{\bar\nu_j\}$), exchange ($\tilde{t}_{ij}^{\bar\mu_j\bar\nu_i}$: $i \rightarrow \{\bar\mu_j\},~j \rightarrow \{\bar\nu_i\}$) and charge transfer ($\tilde{t}_{ij}^{\bar\mu_i\bar\nu_i}$: $ij \rightarrow \{\bar\mu_i,\bar\nu_i\}$ for type 1 and $\tilde{t}_{ij}^{\bar\mu_j\bar\nu_j}$: $ij \rightarrow \{\bar\mu_j, \bar\nu_j\}$ for type 2) joint excitations, respectively. They exhibit different localities with respect to the $ij$ separation.}
For {an diagonal electron pair $ii$} in the same LMOs, there is {only} the vertical correlation in the subspace $\{\Psi^{(\mathrm{vt})}_{ii}\}$.
We map the feature amplitude set $\mathbf{\widetilde{T}}_{ij}^{(\mathrm{vt, ex, ct1, ct2})}=\{\tilde{t}_{ij}^{\bar\mu_i\bar\nu_j},\tilde{t}_{ij}^{\bar\mu_j\bar\nu_i},\tilde{t}_{ij}^{\bar\mu_i\bar\nu_i},\tilde{t}_{ij}^{\bar\mu_j\bar\nu_j}\}$ which is discussed in section 2.3, to these subspaces for each electron pair $ij$ separately,
\begin{eqnarray}
 \tilde{t}_{ij}^{\bar\mu_i\bar\nu_j}  &\mapsto  & \Psi^{(\mathrm{vt})}_{ij},  \\
 \tilde{t}_{ij}^{\bar\mu_j\bar\nu_i}  &\mapsto  & \Psi^{(\mathrm{ex})}_{ij},  \\
 \tilde{t}_{ij}^{\bar\mu_i\bar\nu_i}  &\mapsto  & \Psi^{(\mathrm{ct1})}_{ij},  \\
 \tilde{t}_{ij}^{\bar\mu_j\bar\nu_j}  &\mapsto  & \Psi^{(\mathrm{ct2})}_{ij}.
\end{eqnarray}
Above, $\tilde{t}_{ij}^{\bar\mu_i\bar\nu_j}$, $\tilde{t}_{ij}^{\bar\mu_j\bar\nu_i}$, $\tilde{t}_{ij}^{\bar\mu_i\bar\nu_i}$ and $\tilde{t}_{ij}^{\bar\mu_j\bar\nu_j}$ are the amplitude elements mapping these subspaces to the vertical, exchange, charge transfer electron correlations, respectively.

Here, the feature  vector $\mathbf{\widetilde{T}}_{ij}^{(\mathrm{vt, ex, ct1, ct2})}$ does not solve the post-HF equations for the exact wave function amplitudes, but is rather used {as highly approximated solution} to extract various electron correlation characters from training data. By the proof of principle demonstration, we will show that this simple set of descriptors can deliver transferable learning accuracy and efficiency across chemical compositions and molecular sizes.  {The $\mathbf{\widetilde{T}}^{(\mathrm{vt, ex, ct1, ct2})}_{ij}$ vector is then converted to $\mathbf{\widetilde{T}}^{c(\mathrm{vt, ex, ct1, ct2})}_{ij}$ which is further preprocessed with  $\mathbf{\widetilde{K}}^{(\mathrm{vt, ex, ct1, ct2})}_{ij}=\{\widetilde{K}_{ij}^{\bar\mu_i\bar\nu_j},\widetilde{K}_{ij}^{\bar\mu_j\bar\nu_i},\widetilde{K}_{ij}^{\bar\mu_i\bar\nu_i},\widetilde{K}_{ij}^{\bar\mu_j\bar\nu_j}\}$ integral vector via the following element-wise product,}
\begin{equation}
\tilde{\mathbf{e}}_{ij} = \mathbf{\widetilde{T}}_{ij}^{c(\mathrm{vt, ex, ct1, ct2})}\cdot \mathbf{\widetilde{K}}_{ij}^{(\mathrm{vt, ex, ct1, ct2})}, \label{eij}
\end{equation}
where the pseudo-energy tensor $\tilde{\mathbf{e}}_{ij}$ has the dimension $(4, \tilde{N}_\mathrm{OSV},\tilde{N}_\mathrm{OSV})$ with $\tilde{N}_\mathrm{OSV}$ the number of only a few expressive OSVs (e.g., $\tilde{N}_\mathrm{OSV}=8$). {Here, the $\mathbf{\widetilde{K}}^{(\mathrm{vt, ex, ct1, ct2})}_{ij}$ vector contains the repulsion integrals that are transformed from the half-AO $(i\alpha \vert j\beta)$ via the OSV excitation manifolds of different character, that is, $\widetilde{K}_{ij}^{\bar\mu_i\bar\nu_j}=(i\mu_i \vert j\nu_j)$, $\widetilde{K}_{ij}^{\bar\mu_j\bar\nu_i}=(i\mu_j \vert j\nu_i)$, $\widetilde{K}_{ij}^{\bar\mu_i\bar\nu_i}=(i\mu_i \vert j\nu_i)$ and $\widetilde{K}_{ij}^{\bar\mu_j\bar\nu_j}=(i\mu_j \vert j\nu_j)$ for the vertical, exchange and charge transfer excitations, respectively. For one $ij$ pair, the cost of computing these integrals grows according to $\mathcal{O}(\tilde{N}_\mathrm{OSV}N)$.} The resulting $\tilde{\mathbf{e}}_{ij}$ encodes different physical interactions for various bond types and are subsequently input to the NN  for mapping LMOs to the NN output of pair correlation energies. Although the total correlation energy is not explicitly or implicitly considered for each molecule during the training, as will be shown later, these  features are capable of predicting accurate electron correlations at both MP2 and CCSD levels of theory.

\subsection{Local Feature Amplitudes}

The fundamental physical information for electron correlation must be contained to improve learning data efficiency. We therefore propose to use the feature amplitudes $ \mathbf{\widetilde{T}}^{(\mathrm{vt, ex, ct1, ct2})}_{ij}$ for different physical interactions as follows, in a form analogous to canonical MP2 amplitudes,
\begin{equation}
 \mathbf{\widetilde{T}}_{ij}^{(\mathrm{vt, ex, ct1, ct2})}=\frac{\mathbf{\widetilde{K}}_{ij}^{(\mathrm{vt, ex, ct1, ct2})}}{\epsilon_{ij}}
 \label{amp}
\end{equation}
for the \textit{vertical}, \textit{exchange} and \textit{charge transfer} interactions, respectively. The denominator is the orbital energy difference by $\epsilon_{ij}={f_{ii}+f_{jj}-f_{\bar\mu_{i}\bar\mu_{i}}-f_{\bar\nu_{j}\bar\nu_{j}}}$ with $f_{\bar\mu_i\bar\nu_{j}}$ the elements of virtual-virtual Fock matrix in the OSV basis. The computational cost of $\mathbf{\widetilde{T}}_{ij}^{(\mathrm{vt, ex, ct1, ct2})}$ is proportional to the number of $ij$ pairs by an insignificant constant (e.g., $\tilde{N}_\mathrm{OSV}^2=64)$ and grows linearly with the number of atoms when $\mathbf{\widetilde{T}}_{ij}^{(\mathrm{vt, ex, ct1, ct2})}$ for all long-distance pairs are neglected. Although the resulting amplitudes according to eq~\ref{amp} are extremely poor for any direct post-HF  calculations, they represent the near-sighted limit of the correlation residual equation in the basis of compact LMOs and OSVs,  and contain the locality pattern for accurately learning MP2 and CCSD electron correlations.
For \textit{vertical} and \textit{exchange} interactions, we have also considered the mutual coupling between them by including the coupling matrix $\Delta_{ij}$,
  \begin{eqnarray}
\mathbf{\widetilde{T}}_{ij}^{(\mathrm{vt})} \longrightarrow \mathbf{\widetilde{T}}_{ij}^{(\mathrm{vt})} +  \Delta_{ij}[\mathbf{\widetilde{T}}_{ij}^{(\mathrm{ex})}], & 
\mathbf{\widetilde{T}}_{ij}^{(\mathrm{ex})} \longrightarrow \mathbf{\widetilde{T}}_{ij}^{(\mathrm{ex})} +  \Delta_{ij}[\mathbf{\widetilde{T}}_{ij}^{(\mathrm{vt})}],\label{delta1}
\end{eqnarray}
\begin{equation}
 \Delta_{ij}[\mathbf{\widetilde{T}}_{ij}] =  [\mathbf{S}_{ij}\mathbf{\widetilde{T}}_{ij}\mathbf{F}_{ij}+\mathbf{F}_{ij}\mathbf{\widetilde{T}}_{ij}\mathbf{S}_{ij} -(f_{ii}+f_{jj})\mathbf{S}_{ij}\mathbf{\widetilde{T}}_{ij}\mathbf{S}_{ij}]/\epsilon_{ij} \label{delta2}
\end{equation} 
where $\mathbf{S}_{ij}$ and $\mathbf{F}_{ij}$ are the $\tilde{N}_\mathrm{OSV}$-by-$\tilde{N}_\mathrm{OSV}$  overlap and Fock matrices in the OSV basis for the pair $ij$, respectively. The costs of computing eqs~\ref{delta1} and \ref{delta2} grow linearly with the number of atoms with a small constant prefactor of $\tilde{N}_\mathrm{OSV}^3$.

{The present T-dNN model is significantly simpler than the recent OrbNet\cite{qiao2020orbnet} and OrbNet-Equi\cite{qiao2022informing} variants based on mean-field descriptors for an embedded graph neural network with node and edge embeddings. However,  the T-dNN many-body effect is naturally recovered by directly mapping simple correlated amplitudes to the correlation energy of each pair, which avoids embedding hyperparameters; the set of prioritized local OSVs enables an automatic selection of compact feature space without hyperparameter cutoffs and regularization.}

\subsection{Reference Electronic Structure Calculations with Pair Classification}\label{sec3}

The T-dNN features describe the dynamic electron correlation arising from the virtual excitations from two LMOs. Given that an increasing inter-LMO distance leads to an exponential decrease in associated OSV overlaps due to the locality of LMOs and the compactness of OSVs, the pairwise interaction intensity can be estimated by the ratio between the squared norm of the OSV overlap ($\braket{\bar\mu_{i}|\bar\nu_{j}}$) and the square root of the product of the OSVs numbers ($n_i$ and  $n_j$),\cite{liang2021third}
\begin{equation}
s_{ij} = \frac{\sum_{\bar\mu \bar\nu}\braket{\bar\mu_{i}|\bar\nu_{j}}^2}{\sqrt{n_i n_j}}.
\label{eq:pairscreen}
\end{equation}
The LMO pairs are classified into close, weak, and remote pairs for separate training and predictions according to the magnitude of $s_{ij}$. For close pairs that normally constitute a small fraction in large molecules, the feature contains the pseudo-energy tensor (eq \ref{eij}) for expressing the vertical, exchange and charge-transfer electron correlations. For weak pairs that typically account for a majority of the total pairs, e.g., over 70\% for Nonactin molecule according to eq \ref{eq:pairscreen}, the vertical correlation feature alone is sufficient  to recover the correlation information as demonstrated in our previous work. The remote pairs identified by a tight threshold for $ s_{ij}$ can be discarded to save computational efforts with negligible loss in accuracy. \cite{liang2021third} {Based on the pair classification, a large number of $J_{ij}=\left(ii\vert jj\right)$ and $K_{ij}=\left(ij\vert ij\right)$ integrals arising from unimportant remote pairs can be neglected.}

\begin{table}[ht]
\centering
\caption{Computational costs of all major steps for feature generation with the numbers of occupied LMOs ($O$) and atoms ($N$). The asymptotic costs assume the linear growth of LMO pairs with $N$, the integral sparse-fitting and a small number of OSVs (e.g., 8 OSVs in this work). OSVs are generated by the interpolative decomposition.\cite{liang2021third}}\label{costs}
\begin{tabular}{ccc}
\hline\hline
Steps	&  Asymptotic  costs	&	Asymptotic costs per MO/pair	\\
 \hline
RHF energy	& $\mathcal{O}(N^{4})$	&	\\	
Boys localization	& $\mathcal{O}(N^3)$	&	\\	
$J_{ij}$ and $K_{ij}$ elements &	$\mathcal{O}{(O^2N^2)}$ 	&	$\mathcal{O}{(N^2)}$	\\	
OSV generation	&  $\mathcal{O}(ON)$	     &	$\mathcal{O}(N)$	\\	
OSV $\mathbf{S}_{ij}$ matrix	&  $\mathcal{O}{(O^2N)}$    & $\mathcal{O}(N)$ \\	
OSV $\mathbf{F}_{ij}$ matrix	&  $\mathcal{O}{(N^2)}$    & $\mathcal{O}(N)$ \\	
OSV $\mathbf{\widetilde{K}}_{ij}$ matrix	&  $\mathcal{O}{(N^2)}$    & $\mathcal{O}(N)$ \\
Feature amplitudes $\mathbf{\widetilde{T}}_{ij}$ & $\mathcal{O}(N)$ & -- \\
Pseudo-energy input  $\tilde{\mathbf{e}}_{ij}$ &  $\mathcal{O}(N)$ & -- \\
\hline\hline
\end{tabular}
\end{table}

{Assuming the rapid attenuation of LMO pair interactions, the analysis of asymptotic computation costs in main steps is given in  Table~\ref{costs}. The major cost in the feature preparation lies in the baseline Hartree-Fock calculations which scale as $\mathcal{O}(N^4)$ with the number of atoms, accounting for 90\% of the computing time.  The computations of the feature set and other intermediate quantities have been fully parallelized based on the MPI framework that is described in  our previous work.\cite{liang2021third} By distributing strong and weak pairs over all processors, the computation in many steps has the linear scaling cost per each task. The $\mathbf{S}_{ij}$ matrices for all pairs are needed for pair classification but the cost is very minor compared to other steps. The $\mathbf{J}_{ij}$ and $\mathbf{K}_{ij}$ costs are dominated by the localization transformation from canonical density fitting integrals for which the pair screening and sparse fitting were not applied. The $J_{ij}$ and $K_{ij}$ generation appears most expensive at the asymptotic scaling of $\mathcal{O}(N^2)$ per LMO pair.  However, in the ML context, we would anticipate a linear scaling cost for $J_{ij}$ and $K_{ij}$ computations, if it is possible to discard more atomic and fitting shells when  the numerical rigor of these integrals can be sacrificed without impairing their expressive feature.}

The T-dNN model was built by training OSV-CCSD \cite{yang2012orbital} pair correlation energies obtained by the MOLPRO 2020.1 quantum chemistry package.\cite{werner2020molpro} The atomic basis functions and Foster-Boys LMOs\cite{foster1960canonical} were extracted from MOLPRO and fed into our in-house OSV-MP2 program for feature generations. An OSV selection threshold of $10^{-6}$ was used for evaluating the reference OSV-CCSD energies, which yielded errors within $\pm$0.2 kcal/mol in relative energies compared to canonical CCSD values for several molecules including $n$-Butane, $n$-Pentane, and $n$-Hexane conformers.\cite{gruzman2009performance} The OSV-CCSD pair energies were categorized following the metrics (eq \ref{eq:pairscreen}) for OSV-MP2 feature calculations.  Density fitting\cite{weigend1998ri} and frozen core approximation were employed for both feature and reference energy calculations. 

\subsection{Training Deep Neural Network}\label{sec4}

The fully connected multilayer feed-forward NN architecture has been employed for training molecular data. Our T-dNN models were built with Keras \cite{chollet2015keras},  a Python interface to the TensorFlow backend.  Three separate NNs have been utilized  to provide universal approximators with sufficient hidden neurons \cite{Hornik1989MultilayerFN} for constructing and refining the mapping between the compact electronic descriptors and pair correlation energies for all diagonal, close off-diagonal and other selected off-diagonal pairs, respectively. Prior to inputting the pseudo-energy tensor into the next NN layer, this feature tensor is further symmetrized, merely for numerically stabilizing the input for each orbital pair. {The symmetrization procedure takes a symmetric combination via $\frac{\tilde{\mathbf{e}}_{ij}+\tilde{\mathbf{e}}_{ji}}{2}$ for the diagonal and upper-triangular blocks of the pseudo-energy input matrix for each pair $ij$, patched with its lower-triangular block using an antisymmetric combination via $-\frac{\vert\tilde{\mathbf{e}}_{ij}-\tilde{\mathbf{e}}_{ji}\vert}{2}$ with the enforced negative sign for removing the numerical instability.}
The pseudo-energy tensor is further supplemented by the occupied blocks of Fock ($f_{ij}$), Coulomb ($J_{ij}$) and exchange ($K_{ij}$) elements. As we separately train the NNs for all diagonal, close off-diagonal and other selected off-diagonal pairs, we do not customize the loss function, although the pair correlation energies differ by orders of magnitude. Otherwise, the training process would be strongly biased towards the close pairs as their contributions are much larger than those from weaker pairs, the latter of which can be added up however to a substantial contribution. We therefore chose the mean absolute error as the loss function, giving equal sample weights to all training pair energies. As the pseudo-energy tensor is non-redundant with respect to the training pairs, no feature selection has been applied for preventing the overfitting issue. For the small network scale we have exploited, the use of regularization is unnecessary.

The predicted total correlation energy for a molecule was computed by adding the individual prediction for pair correlation energies according to Nesbet’s theorem. We have used scikit-learn \cite{scikit-learn} for data preprocessing, and standardized quantities for training all selected pairs belonging to a molecule, except for the hidden neurons for which no batch normalization was applied. The rectified linear activation function (ReLU) \cite{agarap2018deeprelu} has been used for all hidden neurons except for the final output which is just a linear activation to collect the pair energy. Exhaustive hyperparameter tuning was not yet attempted in the current work. The Adam optimizer\cite{kingma2014adam} has been found to yield better performance for the optimization of the NN parameters than other gradient algorithms, with a learning rate of 0.001, and decay of 0.0001. We have found that 3-5 hidden layers and 10-20 hidden neurons for each layer are satisfactory for short alkanes. For more complex organic molecules and water clusters, 6-10 hidden layers with 100-200 neurons each layer were used for training. About 2000 epochs or fewer have been applied without early-stopping.

\section{Results}\label{sec5}

\subsection{Straight-chain, branched and cycloalkanes}

\begin{figure}[H]
\includegraphics[width=10cm]{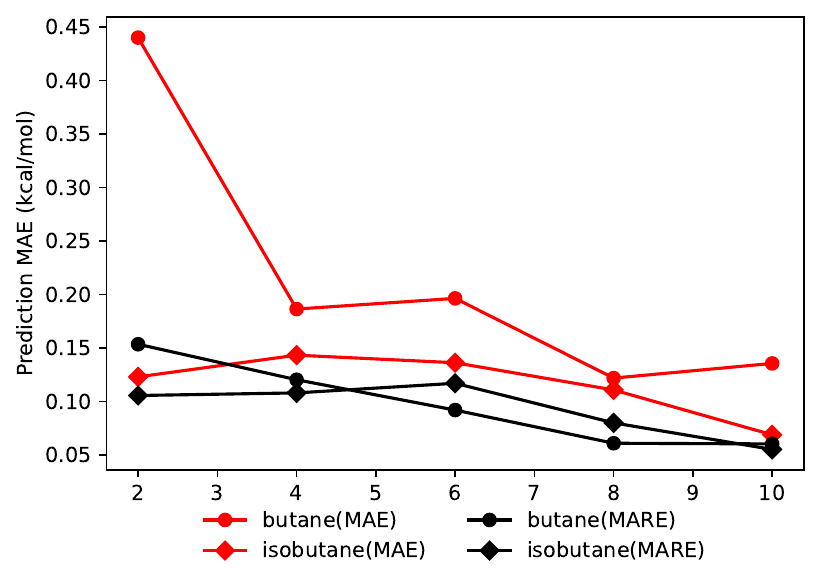}
\centering
\caption{Prediction CCSD/cc-pVTZ errors for $n$-butane and isobutane molecules with respect to the number of OSVs.}
\label{fig:osv}
\end{figure}

We first demonstrate the prediction accuracy and transferability of T-dNN model for a series of straight-chain, branched and cycloalkanes by training only much shorter ethane (C$_2$H$_6$) and propane (C$_3$H$_8$) molecules. The reference geometries of the chain and branched alkanes are obtained from MOB-ML dataset.~\cite{cheng_welborn_christensen_miller_2019} Figure~\ref{fig:osv} shows that the mean absolute errors (MAEs) of predicted correlation energies for $n$-hexane and isobutane are consistently below 0.5 kcal/mol for a range of OSV numbers, as compared to the reference CCSD/cc-pVTZ results, and further drop below 0.2 kcal/mol by increasing the number of OSVs in which more feature amplitudes become available. For better cost-accuracy balance, We have therefore fixed $\tilde{N}_\mathrm{OSV}=8$ for all following calculations. The remaining systematic errors, which are given as the mean absolute relative errors (MAREs) measured by shifting the predicted correlation energies with the magnitude of the mean signed error, are only around 0.06 kcal/mol for $\tilde{N}_\mathrm{OSV}=8$.  

\begin{figure}[H]
\includegraphics[width=10cm]{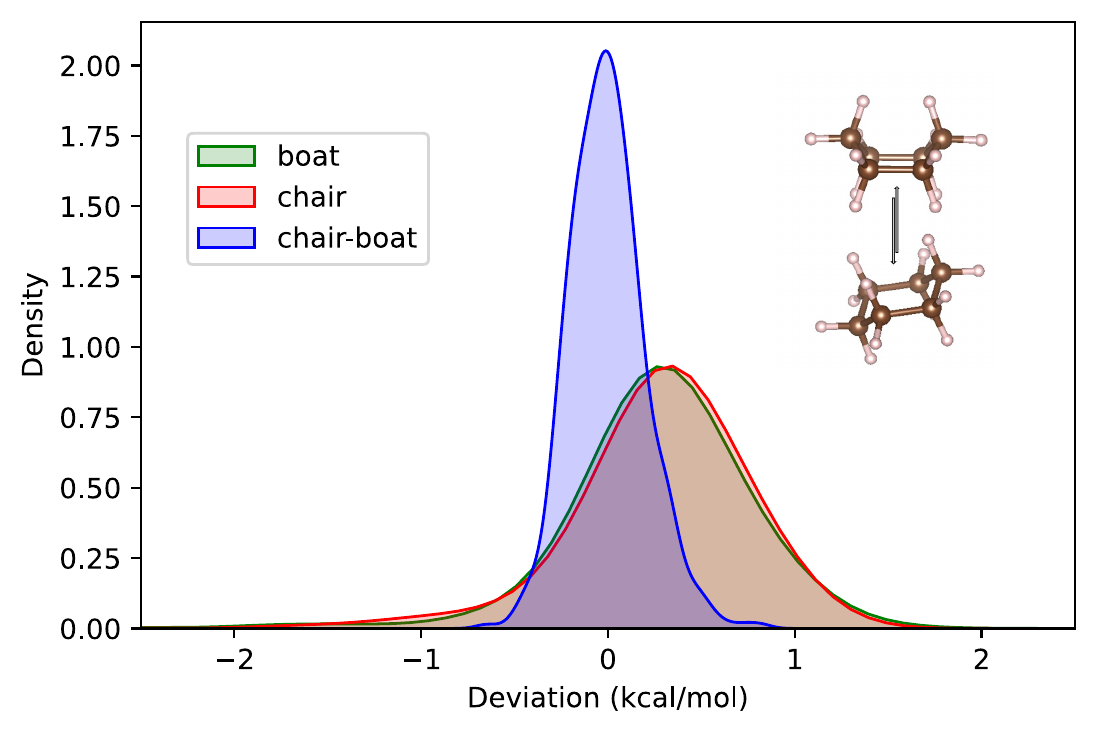}
\centering
\caption{Kernel density estimation (KDE) of the predicted deviations from CCSD/cc-pVTZ reference results for cyclohexane chair (red) and boat (green) energies as well as the chair-boat  energy differences (blue).}
\label{fig:cyclo}
\end{figure}

\begin{figure}[H]
\includegraphics[width=10cm]{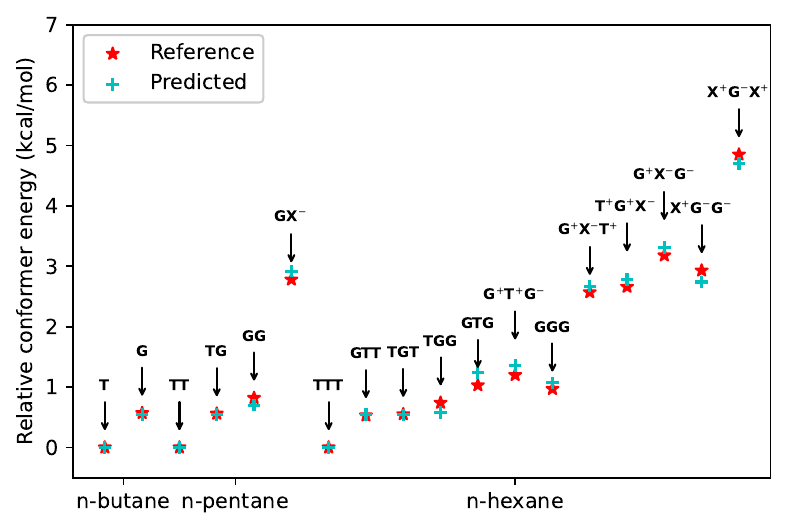}
\centering
\caption{Relative CCSD/cc-pVTZ conformation energies for $n$-butane, $n$-pentane and $n$-hexane molecules with respect to lowest T, TT and TTT conformers, respectively, using the T-dNN model trained on ethane and propane molecules.}
\label{fig:ACONF}
\end{figure}

Moreover, we further trained 491 independent ethane and propane conformers to predict the chair and boat conformations of cyclohexane. The kernel density estimation (KDE) for the boat, chair and conformation energy difference of chair and boat conformation is shown in Figure \ref{fig:cyclo}. Despite the presence of the systematic errors, the two KDE curves of boat and chair conformers are entirely overlapped, and thus the blue KDE curve for the relative conformation energy is symmetrically distributed within $\pm1$ kcal/mol. The average MAE in the conformation energy is only 0.16 kcal/mol (Table S1), with the least error of 0.01 kcal/mol. 
Finally, the relative conformation energies of 18 straight-chain alkane conformations~\cite{gruzman2009performance}(Figure \ref{fig:ACONF}) are accurately predicted by training small thermalized ethane and propane molecules taken from the MOB-ML dataset.~\cite{cheng_welborn_christensen_miller_2019} Although these alkane conformational geometries are very different,  the predicted conformation energies remain chemically accurate on par with CCSD/cc-pVTZ calculations, with the maximal deviation of 0.21 kcal/mol for the GTG conformation of $n$-pentane.

\subsection{QM9 prediction and transferability across molecular datasets}
\begin{figure}[ht]
\includegraphics[width=15cm]{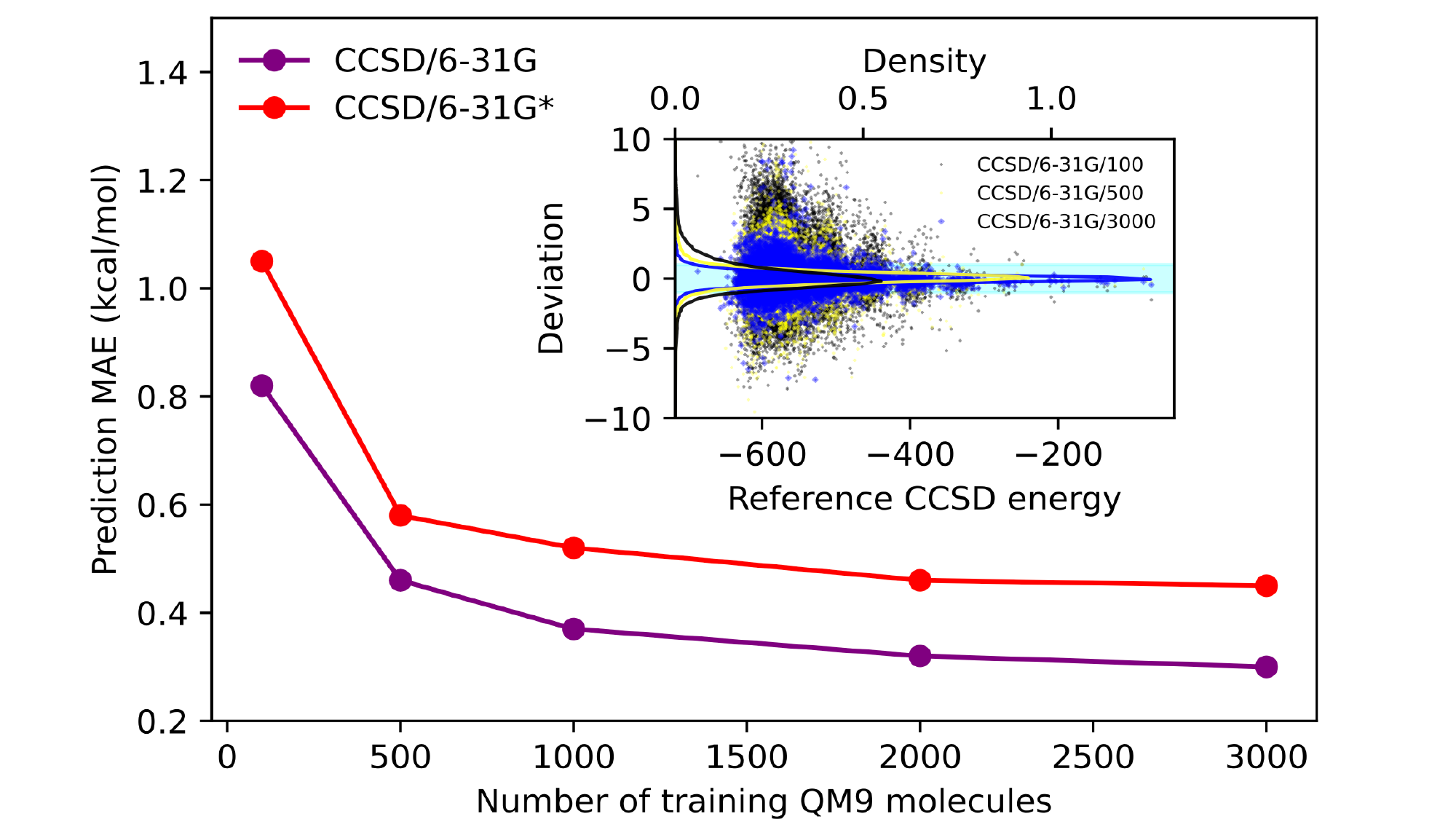}
\centering
\caption{Prediction MAE on the remaining QM9 molecules with different numbers of training molecules. The inset plots the prediction deviations (kcal/mol) from the exact energy (kcal/mol, bottom axis) and the KDE of the deviations (density, top axis) with respect to the T-dNN models trained on 100 (black), 500 (yellow) and 3000 (blue) molecules. The filled area in cyan indicates the error region within $\pm 1$ kcal/mol.}
\label{fig:qm9_curve}
\end{figure}

We applied the T-dNN model to learn correlation energies at the CCSD/6-31G level for QM9 molecular set which encompasses 133885 organic molecules with up to nine C, O, N and F atoms\cite{ramakrishnan2014quantum}. The training sizes of 100 (0.075\%), 500 (0.37\%), 1000 (0.75\%), 2000 (1.5\%) and 3000 (2.2\%) molecules are considered respectively. It can be seen in Figure \ref{fig:qm9_curve} that with only 100 training molecules, the prediction MAE falls already within chemical accuracy for the remaining 99.925\% QM9 molecules, and is gradually reduced to 0.30 kcal/mol with 3000 training molecules.  The KDE distributions for the diagonal and off-diagonal pair energy deviations (Figures S1 and S2) are significantly narrowed with more training data. The symmetric KDE landscape also favors a potential error cancellation.

A recent MP2-based model for learning CCSD correlation energies needs 13392 (10\%) training molecules in QM9 set to obtain similar accuracy~\cite{townsend2020transferable} to the same basis set. Our T-dNN model does not require an MP2 calculation and needs considerably fewer two-electron integrals than what were reported.~\cite{townsend2020transferable, cheng2019universal, husch2021improved} The T-dNN training and prediction is also studied with a larger 6-31G* basis set, and only a slight decrease in accuracy is observed. Given a small amount of data for training, we anticipate that the use of more training data, together with an improved descriptor ans{\"a}tz (e.g., an inclusion of the angular dependence between orbital pairs), may further enhance the T-dNN accuracy.

\begin{figure}[H]
\includegraphics[width=15cm]{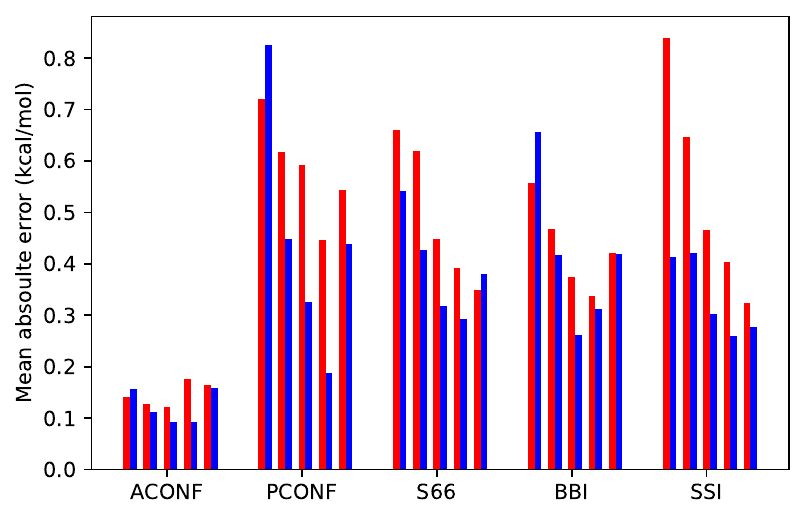}
\centering
\caption{Prediction MAE on the CCSD relative conformation energies for ACONF and PCONF datasets or interaction energies for S66, BBI and SSI datasets. Within each dataset, the bars from the left to right represent training 100, 500, 1000, 2000 and 3000 QM9 molecules, respectively. The MAEs with 6-31G (6-31G*) basis sets are given in blue (red).}
\label{fig:small}
\end{figure}

By training molecular monomers in QM9 dataset, we further assess the transferability of the resulting T-dNN model for predicting molecules in other datasets, including biomolecular interactions. The conformation energies of 18 alkanes in the ACONF~\cite{gruzman2009performance} dataset, 11 tripeptides in PCONF~\cite{vreha2005structure} dataset, as well as 66, 100 and 3380 non-covalent interaction energies between the biomolecular dimer complexes in S66~\cite{rezac2011s66}, BBI and SSI~\cite{burns2017biofragment} datasets are predicted. In Figure~\ref{fig:small}, it can be seen that the MAE is still less than 1.0 kcal/mol for all aforementioned datasets by training only 100 QM9 molecules. The minimal error is observed for the ACONF dataset with only about 0.1 kcal/mol, which is consistent with the excellent performance on alkanes described above. Although the T-dNN model is not trained on the target datasets, the prediction error in general further decreases with increasing the number of the training QM9 molecules. This suggests that the underlying learning is not system specific and can be applied to predict molecules of diverse nature that are not well represented in the training set. The non-covalent interactions are also well reproduced, including the peptide backbone-backbone interactions in the BBI dataset and the sidechain-sidechain interactions in the SSI dataset, similar to the prediction accuracy for monomers in the QM9 dataset. This is again another strong evidence for a transferable mapping that the T-dNN model constructs. To reach even better accuracy, solely increasing the number of QM9 molecules might not help. Instead, one can combine a small amount of the molecules or complexes in the target datasets with QM9 molecules together for training, or alternatively, one can use a pre-trained T-dNN model with QM9 molecules and refine it by transfer learning  upon various application needs.

{We have further trained the T-dNN model on 2000 randomly selected QM9 molecules at the MP2/cc-pVTZ level of theory, and predicted the peptide backbone-backbone in the BBI dataset, sidechain-sidechain interactions in the SSI dataset, various non-covalent interactions in the S66 dataset, and finally the S22 dataset. The MAEs obtained are 0.54, 0.49, 0.59 and 0.84 kcal/mol for these datasets, respectively, which shows that the T-dNN model can make chemically accurate prediction of non-covalent interactions by training smaller molecules dominated by covalent interactions. As the increasing basis set size somewhat smears the electronic locality with diffuse functions, a moderately larger dataset size is necessary than those with compact basis sets. The MAE on S22 dataset is notably larger than other datasets by about 0.3 kcal/mol. The error distribution reveals that the largest errors are from the complexes dominated by $\pi$-$\pi$ dispersion interactions. The largest errors result from the perpendicular ethene dimer ($9^{th}$ complex) and ethene$\cdots$ethyne dimer ($16^{th}$ complex) with perpendicular $\pi$ bonds. However, the errors are less severe for stacked complexes with aromatic-aromatic interactions ($12^{th}$-$14^{th}$ complexes). Surprisingly, the stacked DNA base pair A$\cdots$T ($15^{th}$ complex) has only a prediction deviation of 0.14 kcal/mol. Finally, we note that the recent OrbNet-Equi/SDC21 model\cite{qiao2022informing}, which was trained with 235834 samples by explicitly including sidechain dimers from BFDb-SSI and JSCH-2005\cite{jurevcka2006benchmark} datasets, has achieved an MAE of 0.35 kcal/mol for assessing S66X10 \cite{smith2016revised} non-covalent interactions of the equilibrium geometries with respect to the DFT/def2-TZVP reference data.}

\subsection{Comparison between ML models}

\begin{figure}[H]
\includegraphics[width=15cm]{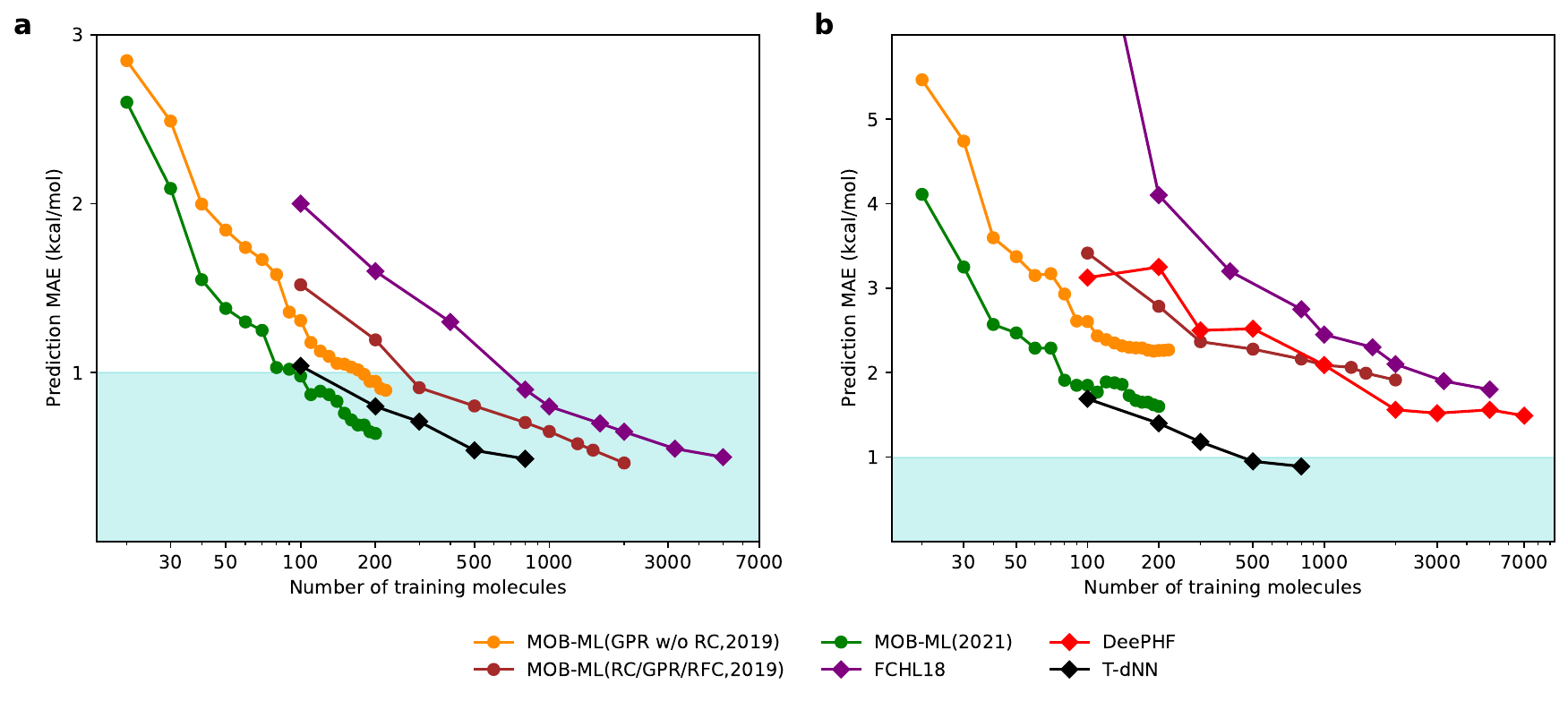}
\centering
\caption{Comparison of T-dNN on the thermalized QM7b-T dataset containing up to 7 heavy atoms and GDB-13-T dataset with up to 13 heavy atoms with reported models. a. Prediction MAE on the remaining QM7b-T molecules with different numbers of training QM7b-T molecules. b. Prediction MAE on the GDB-13-T dataset with different numbers of training QM7b-T molecules.}
\label{fig:qm7btogdb13}
\end{figure}

Here, we dive deeper into different aspects of the T-dNN transferability by comparing it with several recently reported models at the MP2/cc-pVTZ level of theory. We first investigated the transferability of T-dNN within the QM7b-T dataset,~\cite{cheng_welborn_christensen_miller_2019} which is a thermalized dataset containing 7211 organic molecules with at most seven C, O, N, S, and Cl heavy atoms commonly found in drug-like compounds. Figure \ref{fig:qm7btogdb13}a presents the comparison of the prediction performance on the remaining QM7b-T molecules between different ML models with respect to  the training size of QM7b-T molecules. The present T-dNN achieves chemical accuracy within an MAE of 1.04 kcal/mol with 100 training molecules. In contrast, MOB-ML (GPR w/o RC, 2019)~\cite{cheng2019universal}  and MOB-ML (RC/GPR/RFC, 2019)~\cite{cheng2019regression} used approximately 150 and 250 molecules to arrive at similar accuracy, respectively. Moreover, the improved MOB-ML (GPR w/o RC, 2021) model has achieved 0.98 kcal/mol with 100 training molecules. Among all, the FCHL18 model that features atomic descriptors~\cite{faber2018alchemical} appears to need about 700 training  molecules to reach chemical accuracy, which is significantly more than the models based on molecular orbitals. By the increasing number of training molecules, all models make predictions eventually with continually improved accuracy, albeit at different improvement rates. T-dNN has 0.49 kcal/mol MAE with 800 training molecules, and is competitive to MOB-ML variants, while FCHL18 requires 5000 training molecules for the desired accuracy. An improved MOB-ML (GMM/GPR, 2022) model~\cite{cheng2022UC} with unsupervised clustering yields a slightly better MAE of about 0.35 kcal/mol. This improved performance of MOB-ML (GMM/GPR, 2022) can be readily attributed to better learning curves for each cluster in which the orbital pairs are implicitly grouped prior to regression,  and the distribution within each cluster becomes simpler. Similarly, the T-dNN orbital pairs may also be naturally clustered into different bond types by extracting the interaction patterns in the pseudo-energy tensor (eq~\ref{eij}), which will be exploited in our future work. It will be clear later that T-dNN model does not overfit towards QM7b-T dataset and can be generalized to other datasets consisting of larger molecules with competitive accuracy.

The ability of accurately learning electron correlations in larger molecules from small molecules is highly desirable since large molecules are much more computationally expensive and sometimes practically infeasible at a high level of theory. To illustrate this for T-dNN, we examined the transferability of the model by solely training QM7b-T molecules and applied the trained model to GDB-13-T,~\cite{cheng_welborn_christensen_miller_2019} a thermalized dataset of 6000 conformations of organic molecules with 13 heavy atoms consisting of C, O, N, S, and Cl. As seen in Figure \ref{fig:qm7btogdb13}b, T-dNN yields an MAE of 1.69 kcal/mol with 100 training molecules, slightly better than MOB-ML (GPR w/o RC, 2021) of 1.85 kcal/mol.~\cite{husch2021improved} Moreover, T-dNN gradually improves the GDB-13-T prediction accuracy by using more QM7b-T training molecules and achieved 0.89 kcal/mol MAE with 800 training molecules. Using 6500 QM7b-T molecules (90\%) for training, the recent MOB-ML (GMM/GPR, 2022) model with unsupervised clustering~\cite{cheng2022UC} obtains an appealing MAE of 0.16 kcal/mol for the prediction on the rest of 711 QM7b-T molecules, and an MAE of 0.86 kcal/mol for the prediction of GDB-13-T molecules. As a sharp contrast, the DeePHF~\cite{chen2020ground} and FCHL18 model\cite{faber2018alchemical} using atomic feature have the MAEs of about 1.49 kcal/mol and 1.80 kcal/mol with 7000 and 5000 training molecules, respectively. The partition of correlation energy into orbital pairwise contributions for T-dNN and MOB-ML has a rigorous theoretical foundation based on Nesbet’s theorem. Additionally, the T-dNN descriptors are further motivated from local correlation feature expressed in compact OSVs for learning energy components of different characters. On the other hand, the partition into individual atomic contributions might have hampered the QM physical representation which requires more data. This comparison reveals that T-dNN is a promising model for accurate low-data learning and prediction, with energy transferability to scale up for larger molecules by retaining the prediction accuracy similar to the training quality, without overfitting issues in a smaller dataset such as QM7b-T.

\subsection{Water cluster trajectories and morphologies}

We have demonstrated that T-dNN is capable of delivering an accurate and transferable representation for electron correlations among alkanes and organic systems of various chemical compositions and sizes, as well as some biomolecular non-covalent interactions present in peptides. As a final demonstration, we evaluated the T-dNN transferability among the water clusters, and exploited to what extent the T-dNN prediction remains chemically accurate as well as the  impact of different training cluster sizes. All spherical water cluster geometries were extracted from molecular dynamics (MD) trajectories of \textit{ab-initio} NVT molecular dynamics at the B3LYP/6-31G level of theory corresponding to 350 K maintained by the Langevin thermostat. The MD sampling results in 1000 (H$_2$O)$_8$, 1000 (H$_2$O)$_{16}$, 1000 (H$_2$O)$_{32}$, 260 (H$_2$O)$_{64}$ and 50 (H$_2$O)$_{128}$ water clusters. We trained the T-dNN models on the reference MP2/cc-pVTZ correlation energies for 800 (H$_2$O)$_8$ and 800 (H$_2$O)$_{16}$ clusters separately, and evaluated the errors for the remaining water clusters of the same size that do not belong to the training set as well as other smaller and larger water clusters.

\begin{figure}[H]
\includegraphics[width=12cm]{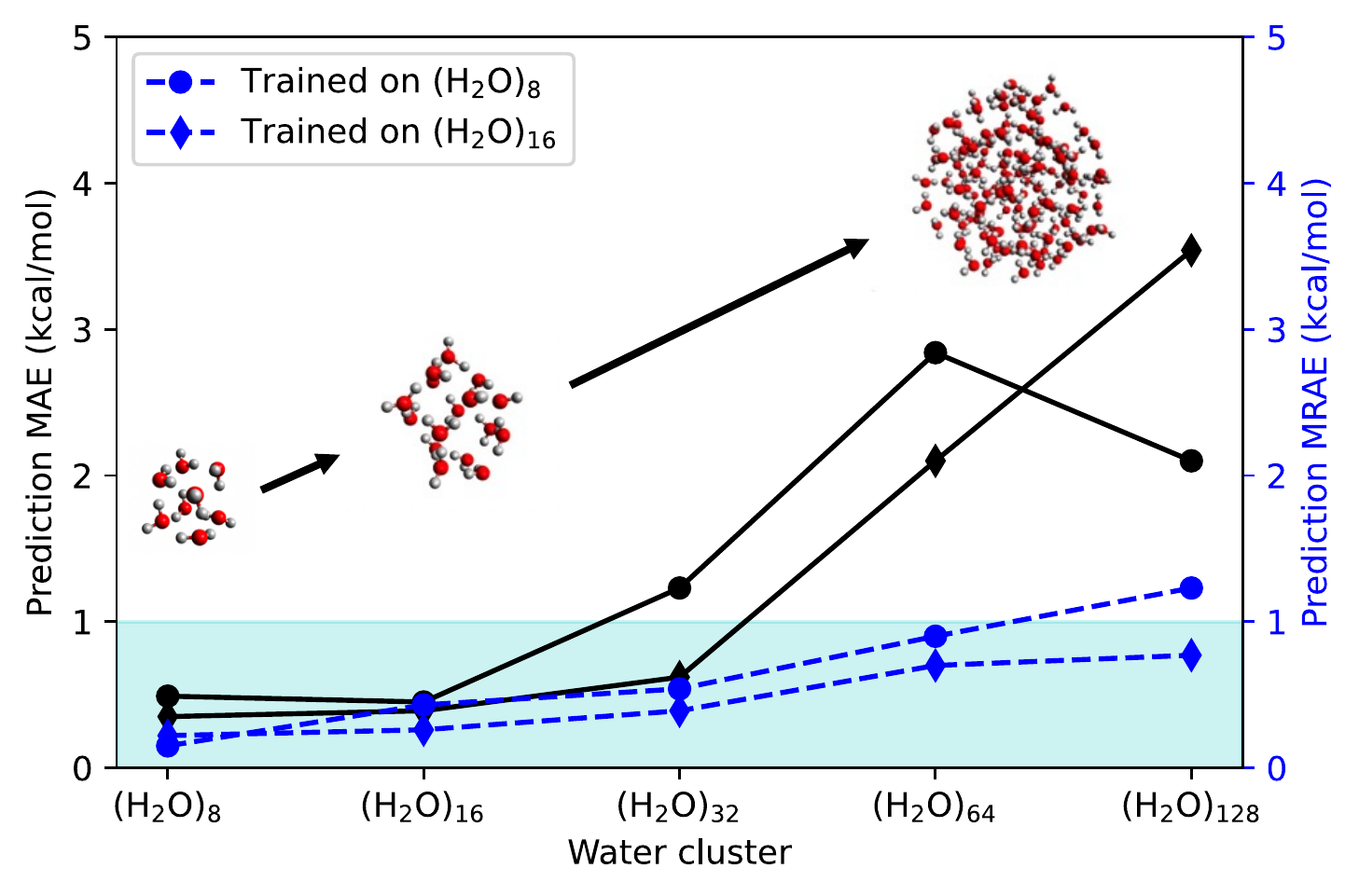}
\centering
\caption{T-dNN prediction errors for spherical water clusters of different sizes sampled from the MD/NVT trajectories.}
\label{fig:water_trans1}
\end{figure}

\begin{figure}[H]
\includegraphics[width=12cm]{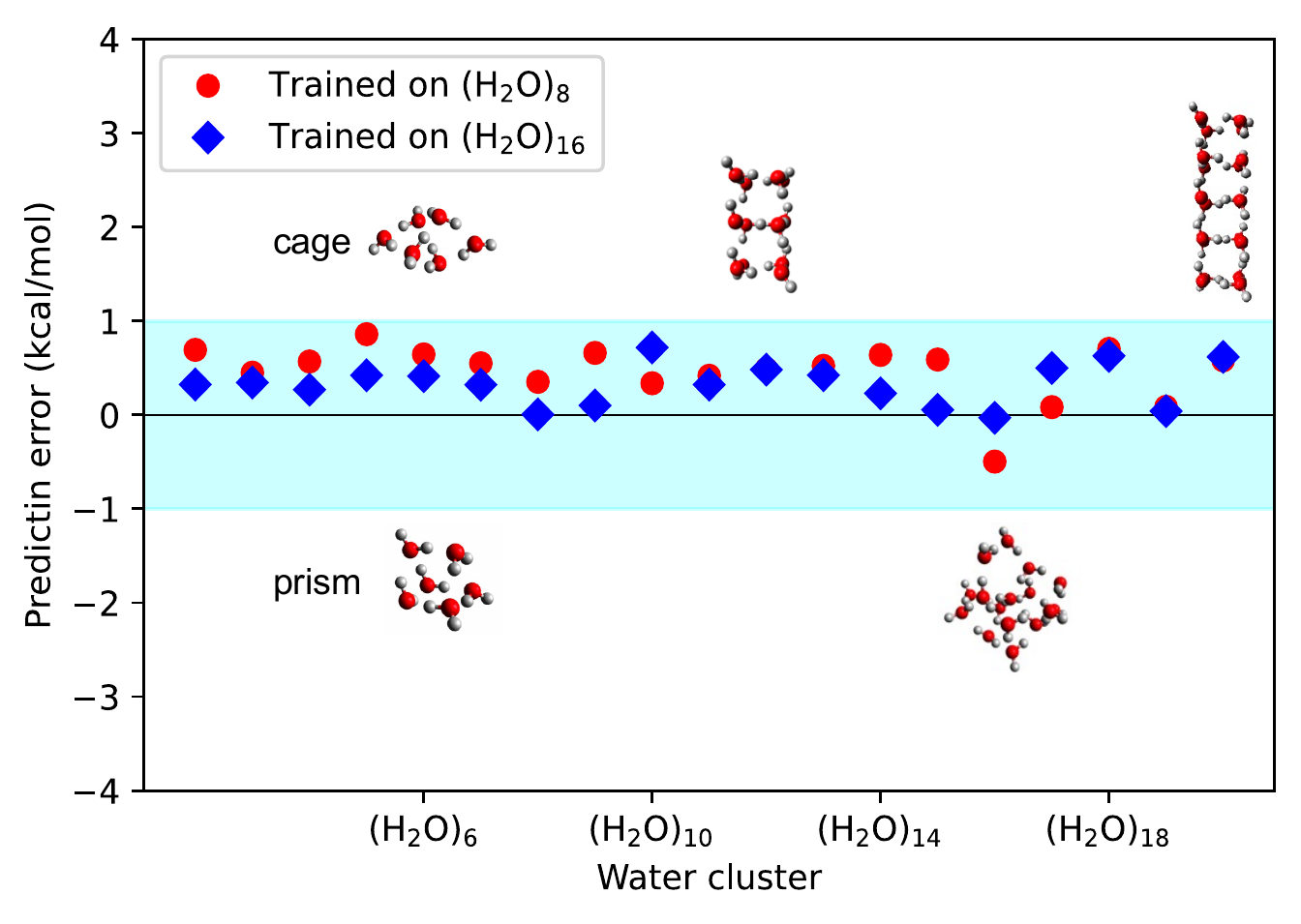}
\centering
\caption{T-dNN prediction errors for water clusters of different sizes and morphologies from the Cambridge Cluster Database.}
\label{fig:cambridge1}
\end{figure}

As seen in Figure \ref{fig:water_trans1}, the T-dNN models separately trained on 800 (H$_2$O)$_8$ and 800 (H$_2$O)$_{16}$ predict the accurate correlation energy of (H$_2$O)$_{16}$ and (H$_2$O)$_{32}$, respectively, within an MAE of 1 kcal/mol. For predictions of larger water clusters up to (H$_2$O)$_{128}$, the total error remains at an accuracy of a few kcal/mol, which grows from (H$_2$O)$_{32}$ to (H$_2$O)$_{128}$ with the number of water molecules.   Surprisingly, the error per 8 H$_2$O molecules actually decreases (Tables S8 and S9), which indicates that the T-dNN model can transfer the underlying mapping between electron pairs and correlations towards larger water clusters. Furthermore, a close analysis of the T-dNN predictions on larger water clusters reveals that the aforementioned errors are virtually systematic. For instance, in Figure S3,  the predicted (H$_2$O)$_{128}$ correlation energies with T-dNN model trained on 800 (H$_2$O)$_{16}$, exhibit an apparent parallel shift relative to the reference MP2/cc-pVTZ correlation energies. This type of errors is fictitious as the potential energy surface curvature is unaffected by a constant global shift. To reveal the random error of the T-dNN models, a global shift on the predicted energies was applied, and the resulting MARE errors are essentially near 1 kcal/mol. For a model trained on (H$_2$O)$_{16}$, the resulting prediction MARE for water clusters of different clusters sizes are consistently lower than those from a model trained on (H$_2$O)$_8$ clusters, and the actual error is only 0.77 kcal/mol for (H$_2$O)$_{128}$.

Moreover, we also investigated the T-dNN transferability on 20 water clusters of other morphologies\cite{cam_waterclusters2001} obtained from the Cambridge Cluster Database. In Figure \ref{fig:cambridge1}, the T-dNN models previously trained on spherical (H$_2$O)$_8$ and (H$_2$O)$_{16}$ clusters sampled from MD trajectories make chemically accurate prediction of all non-spherical water clusters. Although there is an apparent positive systematic error, the conformation energy difference for hexamer prism and cage structures is reproduced within 0.03 kcal/mol with respect to the reference. This shows that our T-dNN model can be transferable to both smaller and larger water clusters that have drastically different shapes and are not represented in the training water clusters, which results from the local and compact nature of T-dNN descriptors.

\begin{figure}[H]
\includegraphics[width=15cm]{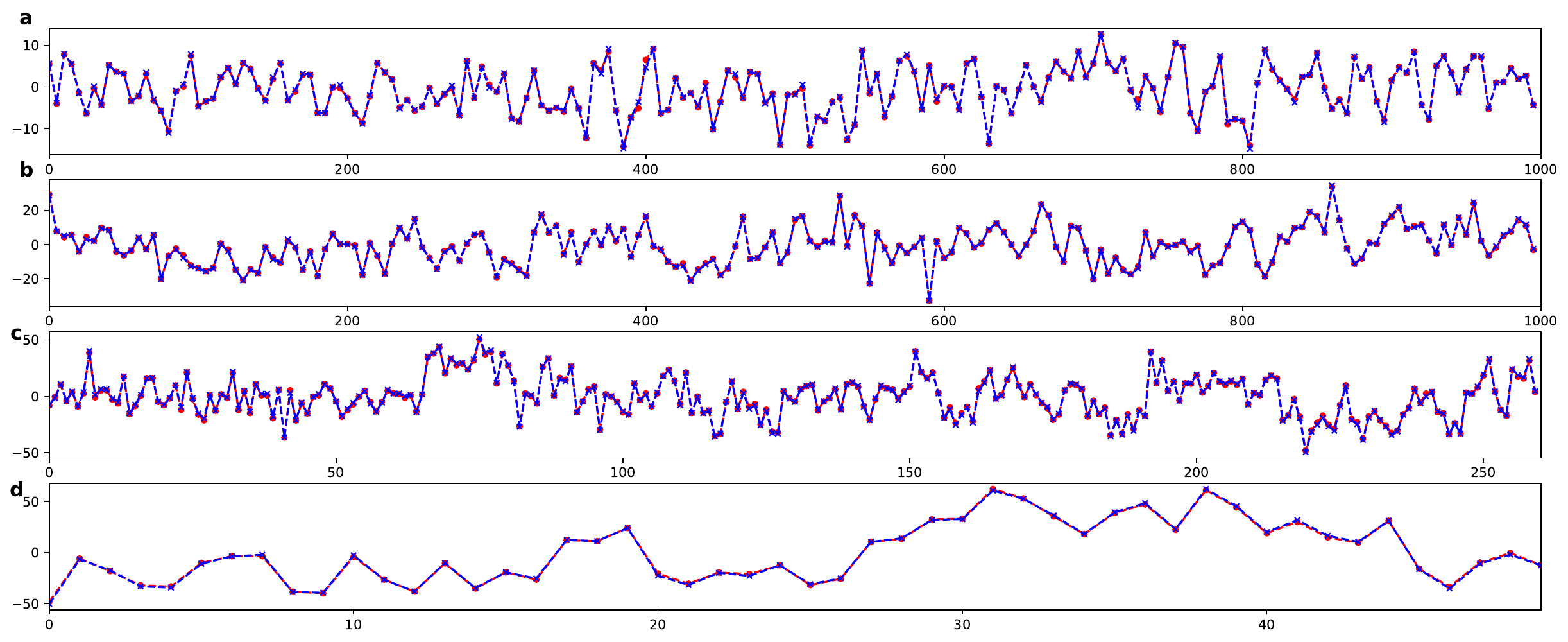}
\centering
\caption{Comparison between the exact MP2-driven MD/NVT energies (red) and the T-dNN predicted trajectory energies (blue) by training 800 (H$_2$O)$_{16}$ for various water clusters of a. (H$_2$O)$_8$, b. (H$_2$O)$_{32}$, c. (H$_2$O)$_{64}$ and d. (H$_2$O)$_{128}$. The systematic errors with the T-dNN predictions at all  simulation time steps were removed by a constant shift of -0.30, 0.56, 2.09 and 3.54 kcal/mol for these water clusters, respectively. The average of the reference energies was set to zero.}
\label{fig:traj}
\end{figure}

Finally, we compare the exact (red) and predicted (blue) MP2 energies for  water clusters of different sizes on their respective trajectories in Figure \ref{fig:traj}(a-d), with a T-dNN model trained on 800 (H$_2$O)$_{16}$. The energy agreement along the trajectories is excellent by removing the systematic error measured by the mean signed error of the predictions. This opens up a possibility to perform \textit{ab-initio} molecular dynamic simulation for large water clusters by training small clusters, and the T-dNN variant for predicting MP2 analytical gradients will be pursued in our future work.

\section{Conclusions}

We have developed a supervised deep NN model which we term T-dNN, based on a quantum chemical feature set from the perspective of local dynamic electron correlations. We have demonstrated that the resulting T-dNN models remain chemically accurate for predicting MP2 and CCSD correlation energies of larger or more complex systems from the training set containing much simpler molecules. Overall, we have demonstrated that the T-dNN approach is both data efficient and highly transferable across alkanes, organic molecules and biomolecular interactions, and water clusters of various sizes and morphologies. More specifically, the utility of the local amplitude decomposition allows an effective learning of many-body electronic interactions of different physical characters.  

We have shown that The T-dNN parameters trained on short ethane and propane molecules can yield high prediction accuracy up to about 0.06 kcal/mol for relative conformation energies of longer branched or cyclic alkanes, including $n$-pentane/$n$-hexane chains, cyclohexane boat and chair conformers.
Moreover, we have demonstrated that the present T-dNN ML model can predict the correlation energies of molecules in the QM9 dataset at chemical accuracy by training as few as 100 molecules, and the prediction error can be further systematically reduced by including more training molecules. For example, The T-dNN model from up to 3000 QM9 training molecules can make accurate energy predictions of molecules across various other datasets, including the BBI and SSI subsets of BFDb database for peptide backbone-backbone interactions and sidechain-sidechain interactions that are not represented in QM9 training molecules. These biomolecular non-covalent interaction energies can be well predicted at similar accuracy to QM9 monomers, even when only QM9 monomers containing covalent bonds are trained. A comparison with other existing models also reveals that T-dNN has exceptional prediction transferability towards molecules of larger size than training molecules. 
We have finally demonstrated the prominent transferability of T-dNN for predicting water clusters containing hydrogen bonds. By training 800 small water clusters, e.g., (H$_2$O)$_{8}$, the T-dNN  prediction exhibits a systematic error that can be removed to yield accurate energies within about 1 kcal/mol for up to spherical (H$_2$O)$_{128}$ as well as the prism and cage (H$_2$O)$_{18}$.
Our results imply that the T-dNN electronic descriptors have intrinsically encoded the many-body feature for capturing various important physical interactions across  both molecular length scales and bonding patterns. 

Our future work will be devoted to linear-scaling T-dNN energy and gradient models by engineering composite atomic/electronic descriptors with sophisticated ML architectures toward large-scale prediction of macromolecular geometries and their molecular dynamics simulations. One important aspect is to further examine the mapping ability from lower-tier MOs than those by Hartree-Fock to alleviate relevant computational costs. We can envisage putting forward cost-effective T-dNN variants for practical application studies on complex many-body chemistry based on learnable knowledge of much smaller molecules.

\begin{acknowledgement}

The authors acknowledge financial supports from the Hong Kong Research Grant Council (17309020), the Hong Kong Quantum AI Lab through AIR@InnoHK program of the Hong Kong Government and the University Postgraduate Scholarship of the University of Hong Kong. Q.L. acknowledges Sijin Li for valuable discussions.

\end{acknowledgement}

\begin{suppinfo}

The file Supporting supporting.pdf contains further results of the computations and is available free of charge.

\end{suppinfo}

\bibliography{manuscript}

\end{document}